\RequirePackage{ifpdf}

\RequirePackage{fix-cm}

\documentclass[%
floatfix,
showkeys,
twocolumn, %
nofootinbib, %
superscriptaddress, %
]{revtex4-1}

\ifpdf
  \usepackage{cmap}
\fi
\usepackage{textcomp}

\usepackage[T1,T2A]{fontenc}

\usepackage[utf8]{inputenx}
\input{ix-utf8enc.dfu}

\usepackage[german,russian,english]{babel}

\usepackage{amsmath}
\usepackage{amssymb}

\usepackage{mathtools}
\mathtoolsset{
showonlyrefs,
mathic = true
}

\allowdisplaybreaks

\usepackage{hyperref}
\hypersetup{backref,
 colorlinks=false}
\hypersetup{pdfborder=0 0 0}

\usepackage{microtype}
\UseMicrotypeSet[protrusion]{alltext}

\usepackage{graphicx}

\usepackage[scanall]{psfrag}

\usepackage{listings}
\usepackage{listingsutf8}
\usepackage[frozencache]{minted}

\usepackage{fixltx2e}

\usepackage{nicefrac}

\makeatletter
\def\ps@pprintTitle{%
     \let\@oddhead\@empty
     \let\@evenhead\@empty
     \let\@oddfoot\@empty
     \let\@evenfoot\@oddfoot}
\makeatother

\usepackage{physics}

\begin{document}

\graphicspath{{image/julia-cas/en/}{image/julia-cas/}{image/}}

  \title{Computer algebra in Julia}

\author{Dmitry S. Kulyabov}
\email{kulyabov-ds@rudn.ru}
\affiliation{Department of Applied Probability and Informatics,\\
  Peoples' Friendship University of Russia (RUDN University),\\
  6 Miklukho-Maklaya St, Moscow, 117198, Russian Federation}
\affiliation{Laboratory of Information Technologies\\
  Joint Institute for Nuclear Research\\
  6 Joliot-Curie, Dubna, Moscow region, 141980, Russian Federation}

\author{Anna V. Korolkova}
\email{korolkova-av@rudn.ru}
\affiliation{Department of Applied Probability and Informatics,\\
  Peoples' Friendship University of Russia (RUDN University),\\
  6 Miklukho-Maklaya St, Moscow, 117198, Russian Federation}

\begin{abstract}
  Recently, the place of the main programming language for scientific
  and engineering computations has been little by little taken by
  Julia. Some users want to work completely within the Julia framework
  as they work within the Python framework. There are libraries for
  Julia that cover the majority of scientific and engineering
  computations demands. The aim of this paper is to combine the usage
  of Julia framework for numerical computations and for symbolic
  computations in mathematical modeling problems. The main functional
  domains determining various variants of the application of computer
  algebra systems are described. In each of these domains, generic
  representatives of computer algebra systems in Julia are
  distinguished. The conclusion is that it is possible (and even
  convenient) to use computer algebra systems within the Julia
  framework.
\end{abstract}


\maketitle

\section{Introduction}
\label{sec:intro}

For mathematical modeling problems, we want to
use the Julia programming
language~\cite{bezanson:2017:julia-numeric-computing,joshi:book:learning-julia,tate:book:seven-more-languages,bezanson:2012:julia-dinamic-language}.

Julia is a problem-oriented language for scientific and engineering
computations. It has a simple syntax.  To facilitate the migration of
researchers and engineers from other languages, Julia borrows good
language constructs from MATLAB, R, Python, and
FORTRAN~\cite{bezanson:2018:julia-dinamism-by-design,auroba:2014:comparison-programming-languages}.. To
support interactive development and optimization of the execution
time, Julia uses JIT compilation. Intrinsically, Julia is a LISP-like
language, and for this reason it seamlessly uses functional
constructs. Julia supports libraries written other in programming
languages.

A reason for choosing Julia is the consistent implementation of the
\emph{single language principle}. The point is that the programming
languages for scientific and engineering problems developed in two
directions---languages for writing high performance programs (FORTRAN,
C, C++) and languages for fast prototyping for nonprogrammers (Matlab,
R, Python).  Julia tries to resolve the \emph{problem of two
  languages}\footnote{The problem of two languages can be described as
  follows. A pro- gram prototype is written in a language that is most
  convenient for the developer. The priority is the speed of
  development. For creating the final program, the code is rewritten
  in a language that can efficiently use the hardware. Here the
  priority is the execution speed. Julia supports both paradigms. That
  is, there is no need to use two languages}.

However, computer algebra problems fall out of the main direction of
Julia development.

Taking into account different kinds of problems, we distinguish the
following domains of using computer algebra systems.
\begin{itemize}
\item Computer algebra systems that provide a large range of
  possibilities and do not require from the user deep knowledge of
  programming. Examples are SymPy, Maxima, Axiom, and Reduce.
\item Computer algebra languages. They can be used to write nontrivial
  programs for symbolic computations. Examples are Wolfram and R-Lisp.
\item Domain-specific computer algebra systems.
\end{itemize}
In this paper, we consider computer algebra systems implemented in
Julia from the viewpoint of these domains of application.

\subsection{Structure of the paper}
\label{sec:structure}

In Section~\ref{sec:sympy}, we consider a computer algebra system for nonprofessional programmers that uses
SymPy library. In Section~\ref{sec:symata}, a computer algebra system based on the principles of Wolfram is considered.
Section~\ref{sec:mtk} is devoted to a problem-oriented symbolic
computations language designed for modeling of
dynamical systems.

  \section{SymPy.lj: interface for the computer algebra system in python}
\label{sec:sympy}

One of the basic advantages of Julia (as well as of
Python) is the possibility of using third party libraries.
For example, Julia is able to use libraries written in C
and FORTRAN. Python has a similar capability
(which is the basis of its popularity). It would be silly
to fail to use these rich opportunities. For calling
Python libraries, Julia uses the package \mintinline{julia}{PyCall}.

The package SymPy (\url{http://sympy.org/}) is the Python library for
symbolic computations. Actually, this is one of the most powerful free
symbolic computations tools. Naturally, it can be used from Julia.

By way of example, consider how the functionality
of SymPy can be used from the Julia environment.
Install \mintinline{julia}{PyCall}, load it, and then import the library
SymPy.
\begin{minted}{julia}
import Pkg
Pkg.add("PyCall")
using PyCall
sympy = pyimport("sympy")
\end{minted}

The further operations are similar to the work in
SymPy. Define the symbolic variable \mintinline{julia}{x} and take the
sine of this variable:
\begin{minted}{julia}
x = sympy.Symbol("x")
y = sympy.sin(x)
\end{minted}

Now, we calculate $\sin(\pi)$:
\begin{minted}{julia}
y.subs(x, pi)
\end{minted}
The result is of the type \mintinline{julia}{PyObject}. In order to use
this object, it should be reduced to a numeric typem
e.g., as
\begin{minted}{julia}
y.subs(x, pi) |> float
\end{minted}
Among other things, this example demonstrates
the functional nature of Julia. In this expression, the
operator \mintinline{julia}{|>} specifies the action of the function on the
right (by analogy with the pipeline in Unix). That is,
the expressions \mintinline{julia}{f(x)} and \mintinline{julia}{x |> f} are equivalent.

e expressions f(x) and x |> f are equivalent.
However, the notation above looks somewhat
cumbersome. A lot of additional code that has no con-
tent has to be written. Fortunately, standard Julia
capabilities can be used for the language extensions.
To make SymPy operations look similarly to the other
computations in Julia, the Python library can be called
using the package SymPy.lj~\cite{sympy.lj:site}:
\begin{minted}{julia}
import Pkg
Pkg.add("SymPy")
using SymPy
\end{minted}

  Then, the example above takes the form
\begin{minted}{julia}
x = Sym("x")
y = sin(x)
subs(y, x, pi) |> float
\end{minted}
For SymPy objects, the type Sym is used.

In principle, this could be the end of discussion.
If the reader has experience in using SymPy in
Python, then he or she will be able to work with
SymPy.lj in Julia.

Julia was created in such a way as to facilitate for the people who
earlier used other programming languages for scientific and
engineering computations (e.g., Matlab, R, or FORTRAN) the migration
to Julia.  For this reason, Julia contains a lot of syntactic sugar;
in particular many operations can be performed in different ways. For
example, a symbolic variable can be specified using a constructor
\begin{minted}{julia}
x = Sym("x")
\end{minted}
and using a macro
\begin{minted}{julia}
@syms x y z
\end{minted}
Variables can also be defined using a function that
simulates the work with SymPy from Python:
\begin{minted}{julia}
x, y = symbols("x y", commutative=false)
\end{minted}
Here, in addition to declaring symbolic variables,
their properties are specified.

Note that the mandatory definition of the variable type contradicts
the dynamic nature of Julia. However, in this case we use an external
library and, therefore, we play by someone else’s rules. Here we have
an analogy with Python, which, when working with external libraries,
uses the conventions of those libraries rather than replaces it syntax
by its own.

SymPy.lj gives Julia’s user the possibility to manipulate symbolic
expressions and provides a convenient interface for this
purpose. However (taking into account the performance of modern
computers) no speed optimization is achieved.

Here are some examples of working with SymPy.lj.  In the manipulations
with algebraic expressions, the operations of factorization (\mintinline{julia}{factor})
and expansion (\mintinline{julia}{expand}) are very useful:
\begin{minted}{julia}
x,y  = symbols("x, y")
factor(x^2 - 2x + 1)
\end{minted}

\begin{equation*}
  \left(x - 1\right)^{2}
\end{equation*}

\begin{minted}{julia}
expand((x-1)*(x+1))
\end{minted}
\begin{equation*}
  x^{2} - 1
\end{equation*}

  In symbolic expressions, Julia constructs, e.g.,
  comprehensions\footnote{Сomprehensions provide a generic efficient
    way for constructing arrays. Their syntax resembles the notation
    of set constructors in the Zermelo--Fraenkel axiomatic.} may be
  used:
\begin{minted}{julia}
expand(prod([x-i for i in 1:5]))
\end{minted}
\begin{equation*}
  x^{5} - 15 x^{4} + 85 x^{3} - 225 x^{2} + 274 x - 120
\end{equation*}

  This is possible because SymPy.lj represents sym-
bolic matrices as arrays with the elements of the type
julia \mintinline{julia}{Sym}, e.g., as
\mintinline{julia}{Array{Sym}}\footnote{Naturally, this incurs
  additional overheads.}.

Let us specify the symbolic matrix
\begin{minted}{julia}
x, y = symbols("x y")
m = [x 1; 1 x]
\end{minted}
\begin{equation*}
  \left[ \begin{array}{rr}x&1\\1&x\end{array}\right]
\end{equation*}

Then, we can perform standard operations on
matrices, e.g., multiply them:
\begin{minted}{julia}
m * m
\end{minted}
\begin{equation*}
  \left[ \begin{array}{rr}x^{2} + 1&2 x\\2 x&x^{2} + 1\end{array}\right]
\end{equation*}
\begin{minted}{julia}
m .* m
\end{minted}
\begin{equation*}
  \left[ \begin{array}{rr}x^{2}&1\\1&x^{2}\end{array}\right]
\end{equation*}

The majority of standard operations on matrices
are also extended for working with symbolic values.
By the way, this is an example of the implementation
of multiple dispatch.

The functions \mintinline{julia}{factor} and
\mintinline{julia}{expand} can simplify the given expression. SymPy
has a lot of functions for various simplifications. For example, there
is the general function juliasimplify, which tries to derive the
simplest form of an expression using various heuristics for this
purpose: \mintinline{julia}{simplify}:
\mintinline{julia}{simplify}:
\begin{minted}{julia}
a = (x + x^2)/(x*sin(y)^2 + x*cos(y)^2)
simplify(a)
\end{minted}
\begin{equation*}
  x + 1
\end{equation*}

Equations and systems of equations can be solved
using the function \mintinline{julia}{solve}. For example, let us solve
the equation
\begin{equation}
  x^2 + 3x + 2 = 0
\end{equation}
for $x$:
\begin{minted}{julia}
x  = symbols("x")
solve(x^2 + 3*x + 2, x)
\end{minted}
\begin{equation*}
  \left[ \begin{array}{r}-2\\-1\end{array} \right]
\end{equation*}

For systems of equations, vector notation may be
used:
\begin{minted}{julia}
x,y  = symbols("x y")
eq1 = x + y - 1
eq2 = x - y - 2
solve([eq1, eq2], [x, y])
\end{minted}
\begin{minted}{julia}
Dict{Any,Any} with 2 entries:
  y => -1/2
  x => 3/2
\end{minted}

Finally, there is differentiation and integration.

The function \mintinline{julia}{diff} is used to compute derivatives
of symbolic expressions. One can compute partial and
higher order derivatives. For example, define the functions
\begin{minted}{julia}
x, y = symbols("x y")
f(x) = exp(-x) * sin(x)
g(x, y) = x^2 + 17*x*y^2
\end{minted}
and compute the derivative $\dv{f(x)}{x}$:
\begin{minted}{julia}
 diff(f(x))
\end{minted}
\begin{equation*}
  - e^{- x} \sin{\left(x \right)} + e^{- x}
  \cos{\left(x \right)}
\end{equation*}
Here we omitted the argument with respect to
which the differentiation is performed. However, it
may be indicated explicitly:
\begin{minted}{julia}
diff(f(x), x)
\end{minted}

We can also compute the higher order derivative $\dv[3]{f(x)}{x}$:
\begin{minted}{julia}
diff(f(x), x, 3)
\end{minted}
\begin{equation*}
  2 \left(\sin{\left(x \right)} + \cos{\left(x \right)}\right) e^{- x}
\end{equation*}

and compute the partial derivative $\pdv{g(x,y)}{x}$:
\begin{minted}{julia}
diff(g(x,y), x)
\end{minted}
\begin{equation*}
  2 x + 17 y^{2}
\end{equation*}

Similarly, symbolic integration can be performed.
Thus, we can compute $\int (x^2+ x + 2) \dd{x}$:
\begin{minted}{julia}
x,y = symbols("x y")
integrate(x^2 + x + 2)
\end{minted}
\begin{equation*}
  \frac{x^{3}}{3} + \frac{x^{2}}{2} + 2 x
\end{equation*}
or the double integral $\int \dd{y} \int \dd{x} xy$:
\begin{minted}{julia}
    integrate(x*y, (x, y))
\end{minted}
\begin{equation*}
  \frac{y^{3}}{2}
\end{equation*}
The definite integral
$\int_{0}^{1} x \dd{x}$ can also be computed:
\begin{minted}{julia}
integrate(x^2, (x, 0, 1))
\end{minted}
\begin{equation*}
  \frac{1}{3}
\end{equation*}

Thus, we have a very rich computer algebra system
that can be used directly from the Julia environment.
\section{Symata: a computer algebra language for Julia}
\label{sec:symata}

  Even though SymPy is an excellent computer algebra system, it has a
  small drawback. This is a ready-to-use system---a ``packaged
  product''. It provides a user layer for symbolic computations but
  not a developer layer that makes it possible to write code. For
  development, one can use the language Symata\footnote{In the case of
    Symata, the developer layer language and user layer language are
    identical. Hence, Symata is simultaneously a programming language
    and a computer algebra system.}~\cite{symata.jl:site} for symbolic
  computations. Symata is implemented in Julia.  The design of Symata
  is based on Wolfram~\cite{wolfram:book:elementary-introduction}.

To use this system, the package Symata must be
installed:
\begin{minted}{julia}
import Pkg
Pkg.add("Symata")
\end{minted}

Symata is loaded in the standard way:
\begin{minted}{julia}
using Symata;
\end{minted}

Pay attention to the following property of Symata.
While SymPy.lj was designed with the aim of using it
directly and seamlessly from the Julia environment,
Symata is a domain-specific language written in Julia.
It is fairly different from Julia. For this reason, one has
to explicitly switch between these languages. Switching
to Julia is made using the function
\begin{minted}{julia}
Julia()
\end{minted}

Using the function
\begin{minted}{julia}
isymata()
\end{minted}
one can switch back to the Symata mode.

  Computations in Symata are similar to the work in Mathematica. For
  example, suppose that we want to find the value of cosine:
\begin{minted}{julia}
expr = Cos(pi * x)
\end{minted}
$$ \text{Cos} \!  \left(  \pi  \ x \right) $$

Specify the value of the variable $x$:
\begin{minted}{julia}
x = 1/3
\end{minted}
$$ \dfrac{1}{3} $$

Now, we get the value of the expression \mintinline{julia}{expr}:
\begin{minted}{julia}
expr
\end{minted}
$$ \dfrac{1}{2} $$

Modify the value of $x$:
\begin{minted}{julia}
x = 1/6
\end{minted}
$$ \dfrac{1}{6} $$

Then the value of the expression also changes:
\begin{minted}{julia}
expr
\end{minted}
$$ \dfrac{3^{\frac{1}{2}}}{2} $$

Reset $x$. Then, the numeric value of the expression
is not calculated:
\begin{minted}{julia}
Clear(x)
expr
\end{minted}
$$ \text{Cos} \!  \left(  \pi  \ x \right)  $$

Actually, a Wolfram-like language is implemented
in Julia. There is no need to describe the syntax of
Symata in more detail here. However, it makes sense
to consider some differences of Symata from Wolfram.
The majority of these differences are due to the desire
to make the syntax of Symata closer to the syntax of
Julia. For example, comments are marked by the symbol
\mintinline{julia}{#}, while in Wolfram, the comments look like
\begin{minted}{mathematica}
(* comment *)
\end{minted}

For specifying lists, square brackets 
\begin{minted}{julia}
[a,b,c]
\end{minted}
$$  \left[ a,b,c \right]  $$
rather than curly brackets \mintinline{mathematica}{{ }} are used. The
list elements may be separated by comas or by line break:
\begin{minted}{julia}
[
    a
    c + d
    Expand((x+y)^2)
]
\end{minted}
$$  \left[ a,c + d,x^{2} + 2 \ x \ y + y^{2} \right]  $$

As in Julia, the argument of a function is specified
in parenthesis rather than in square brackets as in Wolfram:
\begin{minted}{julia}
f(x)
\end{minted}
$$ f \!  \left( x \right)  $$

Some functions have infix notation as in Julia.
  
For example, the function \mintinline{julia}{Map(f,list)} in
infix form is written as
\begin{minted}{julia}
f % list
\end{minted}
and the \mintinline{julia}{Apply(x,y)} as
\begin{minted}{julia}
x .%  y
\end{minted}

  For details, see the Symata documentation.
  
We may conclude that, on the basis of Julia (using
its powerful capabilities in creating domain-specific
languages), an extremely convenient language for
computer algebra is implemented. It is of interest to
mention that Symata uses the library SymPy for many
analytical computations.

\section{Computer algebra in Julia for mathematical programming problems}
\label{sec:mtk}

In the preceding sections, we considered the implementation of
general-purpose computer algebra systems in Julia. However, of great
interest are specialized
domain-specific applications of computer algebra.

The package ModelingToolkit.jl~\cite{modelingtoolkit.jl:site} provides a
specialized computer algebra language for mathematical modeling problems. For this purpose, it uses the
metaprogramming capabilities~\cite{lammel:2018:software-languages} of Julia. Note that
the main mode of operation of ModelingToolkit.jl is
the batch mode rather than the interactive one.

Symbolic variables are declared by the macro
\mintinline{julia}{@variables}:
\begin{minted}{julia}
@variables x y
\end{minted}

Then, these variables can be used in symbolic
expressions (in the Julia syntax):
\begin{minted}{julia}
z = x^2 + y
\end{minted}

For modeling continuous dynamical systems, we
need derivatives. In ModelingToolkit.jl, differential
operators are constructed using the macro
\mintinline{julia}{@derivatives}:
\begin{minted}{julia}
@variables t
@derivatives D'~t
\end{minted}
Here, the differential operator $D = \dv{}{t}$ is specified.
The number of primes \mintinline{julia}{'} indicates the order of the
differential operator. We can write the expression
\begin{minted}{julia}
z = t + t^2
D(z)
\end{minted}
At this point, the differential operator does not
compute anything because it is a lazy operator. However, we also can
obtain the result immediately by
using the function
\mintinline{julia}{expand_derivatives}:
\begin{minted}{julia}
expand_derivatives(D(z))
\end{minted}
\begin{equation*}
1 + 2t
\end{equation*}

Since functions in Julia are first-order objects, the
declared symbolic variables are actually functions.
When declaring variables, dependences may be indicated explicitly:
\begin{minted}{julia}
@variables t x(t) y(t)
\end{minted}

This dependence is taken into account in differentiation:
\begin{minted}{julia}
z = x + y*t
expand_derivatives(D(z))
\end{minted}
The last expression is expanded as
\begin{minted}{julia}
derivative(x(t), t) + y(t) + derivative(y(t), t) * t
\end{minted}

By way of example, solve the classical predator--prey problem~\cite{lotka:1925:physical-biology,kulyabov:2018:formalizms-stochastization}:
\begin{equation}
  \label{eq:lotka}
  \begin{dcases}
    \dv{x}{t} = ax - b y x,\\
    \dv{y}{t} = c x y - d y,
  \end{dcases}
\end{equation}
where $x$ is the population of prey, $y$ is the population
of predators, $t$ is time, and $a$, $b$, $c$, and $d$ are the
coefficients describing the interaction between species.

\begin{figure}
  \centering
  \includegraphics[width=\linewidth]{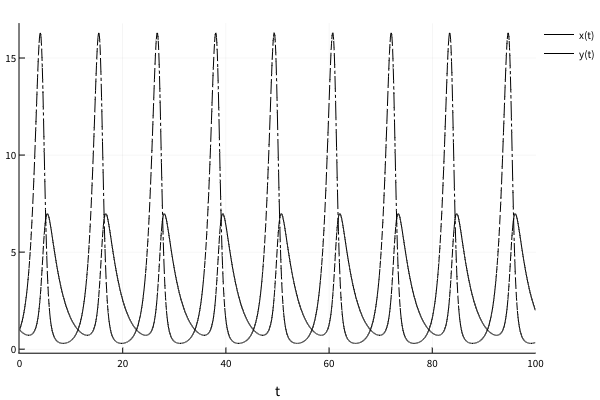}
  \caption{Solution of system~\eqref{eq:lotka}}
  \label{fig:lotka}
\end{figure}

First, load the required packages (if needed):
\begin{minted}{julia}
import Pkg
Pkg.add("ModelingToolkit")
Pkg.add("OrdinaryDiffEq")
\end{minted}

Then, we load the package for plot construction:
\begin{minted}{julia}
Pkg.add("Plots")
\end{minted}
and then link the required packages:
\begin{minted}{julia}
using ModelingToolkit
using OrdinaryDiffEq
using Plots
\end{minted}

Now, we are ready to solve the problem. Declare
variables and differentiation operators. A part of the
variables is declared as parameters.
\begin{minted}{julia}
@parameters t a b c d
@variables x(t) y(t)
@derivatives D'~t
\end{minted}

Now, we simply rewrite system~\eqref{eq:lotka} using the syntax
of Julia:
\begin{minted}{julia}
eqs = [D(x) ~ a *x - b*x*y,
       D(y) ~ c*x*y - d*y]
\end{minted}

By symbolic manipulations, we reduce the system
to the form that is required for the package OrdinaryDiffEq.lj~\cite{rackauckas:2017:differentialequations.jl}:
\begin{minted}{julia}
sys = ODESystem(eqs)
\end{minted}

The further manipulations are performed within
the package OrdinaryDiffEq.lj. Set the initial values of
the variables:
\begin{minted}{julia}
u0 = [x => 1.0
      y => 1.0]
\end{minted}

We also specify the problem parameters:
\begin{minted}{julia}
p  = [a => 1.1
      b => 0.4
      c => 0.1
      d => 0.4]
\end{minted}

We will solve the problem on the time interval
\begin{minted}{julia}
tspan = (0.0,100.0)
\end{minted}

Create the structure describing the entire problem:
\begin{minted}{julia}
prob = ODEProblem(sys,u0,tspan,p; jac=true)
\end{minted}
The additional parameter \mintinline{julia}{jac=true} instructs the
system to symbolically generate the optimized Jacobi
function for improving the work of differential equation solvers.

Finally, we apply the solver available in the package
OrdinaryDiffEq.lj:
\begin{minted}{julia}
sol = solve(prob)
\end{minted}

Now, we plot the solution (see Fig.~\ref{fig:lotka}):
\begin{minted}{julia}
plot(sol)
\end{minted}

This approach fits well into the idea of a specialized
language for scientific and engineering computations.
Under this approach, symbolic computations are performed not by using
a special universal language, but a
special problem-oriented symbolic computation language is created for
each direction. This language
should seamlessly fit with the basic language (Julia in
the case under consideration).
\section{Conclusion}
\label{sec:conclusion}

The (incomplete) survey of computer algebra systems
that can be used with Julia showed that the language itself
and its infrastructure are fairly mature. The computer
algebra systems in Julia have a wide of applications---systems for nonprofessional programmers, (SymPy.lj),
powerful computer algebra languages (Symata.lj), and
domain-specific computer algebra languages (ModelingToolkit.jl). This gives hope that the popularity of
Julia will grow not only in the domain of numerical
computations but also in the domain of symbolic computations.

\begin{acknowledgments}

The publication has been prepared with the support of the ``RUDN University Program 5-100''
and funded by Russian Foundation for Basic Research (RFBR) according to the research project
No~19-01-00645.

\end{acknowledgments}

 \bibliographystyle{elsarticle-num}

\bibliography{bib/julia-cas/cite}

\end{document}


\graphicspath{{image/julia-cas/ru/}{image/julia-cas/}{image/}}

  \title{Компьютерная алгебра на Julia}

\author{Д. С. Кулябов}
\email{kulyabov-ds@rudn.ru}
\affiliation{Кафедра прикладной информатики и теории вероятностей,\\
  Российский университет дружбы народов,\\
  117198, Москва, ул. Миклухо-Маклая, д. 6}
\affiliation{Лаборатория информационных технологий,\\
  Объединённый институт ядерных исследований,\\
  ул. Жолио-Кюри 6, Дубна, Московская область, Россия, 141980}

\author{А. В. Королькова}
\email{korolkova-av@rudn.ru}
\affiliation{Кафедра прикладной информатики и теории вероятностей,\\
  Российский университет дружбы народов,\\
  117198, Москва, ул. Миклухо-Маклая, д. 6}

\begin{abstract}
В последнее время на место основного языка научных и инженерных
расчётов выдвигается язык Julia. У ряда пользователей возникает желание работать полностью
внутри <<экосистемы>> Julia, подобно тому, как происходит работа в
<<экосистеме>> Python. Для Julia существуют библиотеки, покрывающие
большинство потребностей научно-инженерных расчётов. Перед авторами
возникла необходимость использовать символьные вычисления для задач
математического моделирования. Поскольку основным языком реализации
численных алгоритмов мы выбрали язык
Julia, то и задачи компьютерной алгебры хотелось бы решать на
этом же языке. Авторы выделили основные функциональные области,
задающие разные варианты применения систем компьютерной алгебры. В
каждой из областей нами выделены наиболее характерные представители
систем компьютерной алгебры на Julia. В результате авторы делают
вывод, что в рамках <<экосистемы>> Julia возможно (и даже удобно)
использовать системы компьютерной алгебры.
\end{abstract}

  \keywords{Julia,
    SymPy,
    язык Wolfram,
    системы компьютерной алгебра}

\maketitle

\section{Введение}
\label{sec:intro}

Для задач математического моделирования мы ориентированы на
использование языка
Julia~\cite{bezanson:2017:julia-numeric-computing,joshi:book:learning-julia,tate:book:seven-more-languages,bezanson:2012:julia-dinamic-language}.

Язык Julia является проблемно-ориентированным языком для научных и
инженерных расчётов. Язык  имеет простой синтаксис.
Для облегчения перехода научно-инженерных работников, 
Julia использует удачные языковые конструкции, заимствованные из
языков MATLAB, R, Python, FORTRAN~\cite{bezanson:2018:julia-dinamism-by-design,auroba:2014:comparison-programming-languages}. Для поддержки интерактивной
разработки и оптимизации времени исполнения Julia использует
JIT-компиляцию. Внутри Julia является LISP-подобным языком, поэтому
она органично использует функциональные конструкции. Julia
поддерживает библиотеки, написанные на других языках программирования.

Одной из причин выбора Julia является
последовательная реализация \emph{принципа одного языка}. Дело в том,
что в своём развитии языки программирования для научных и инженерных
задач разошлись по двум направлениям: языки для создания
высокопроизводительных программ (FORTRAN, C, C++) и языки для быстрого
прототипирования и использования непрограммистами (Matlab, R,
Python). Julia старается решить \emph{проблему двух
  языков}\footnote{Проблему двух языков можно представить следующим
  образом. Прототип программного комплекса пишется на языке программирования,
  дружественном к разработчику. Приоритет~--- скорость
  написания программного кода. Для создания финального программного
  комплекса код переписывается на языке, позволяющем эффективно
  использовать ресурсы вычислительной техники. Приоритет~--- скорость
  выполнения. Язык Julia поддерживает обе эти парадигмы. То есть для
  него нет необходимости использовать два разных языка программирования.}. Однако
задачи компьютерной алгебры выпадают из основного направления развития
языка Julia.

С учётом различно характера возникающих при этом задач, выделим
следующие области применения систем компьютерной алгебры:
\begin{itemize}
\item Пользовательские системы компьютерной алгебры. Данные системы
  предлагают пользователю широкий спектр возможностей, не требуя от
  него глубоких знаний в программировании. Примеры: SymPy, Maxima,
  Axiom, Reduce.
\item Языки компьютерной алгебры. С помощью них можно писать
  нетривиальные программы символьных вычислений. Примеры: Wolfram,
  R-Lisp.
\item Предметно-ориентированные системы компьютерной алгебры.
\end{itemize}

В работе рассмотрены системы компьютерной алгебры, реализованные на
языке Julia, с точки зрения этих областей применимости.

\subsection{Структура статьи}
\label{sec:structure}

В разделе~\ref{sec:sympy} рассматривается пользовательскую систему
компьютерной алгебры, использующей библиотеку SymPy. В
разделе~\ref{sec:symata} рассматривается язык компьютерной алгебры,
реализованный на основе идеологии языка Wolfram. В
разделе~\ref{sec:mtk} рассмотрен проблемно-ориентированный язык
символьных вычислений для задач непрерывного моделирования
динамических систем.

\section{SymPy.lj --- интерфейс к системе компьютерной алгебры на
  языке Python}
\label{sec:sympy}

Одной из ключевых основ идеологии языка Julia (как и у языка Python)
является возможность использования сторонних библиотек. Например,
Julia может подключать библиотеки, написанные на языках C,
FORTRAN. Сходной идеологией обладает и язык Python (что и послужило
основой его популярности). Поэтому было бы непредусмотрительно не
использовать это богатство. Julia использует для вызова библиотек
языка Python пакет \mintinline{julia}{PyCall}.

Пакет SymPy (\url{http://sympy.org/}) является
библиотекой Python для символьной математики. Фактически, это один из
наиболее мощных свободных пакетов символьных вычислений. Естественно,
что с ним можно работать из Julia.

Рассмотрим для примера, как мы можем использовать функционал SymPy из
среды Julia.
Установим пакет \mintinline{julia}{PyCall}, загрузим его, а затем
импортируем библиотеку SymPy.
\begin{minted}{julia}
import Pkg
Pkg.add("PyCall")
using PyCall
sympy = pyimport("sympy")
\end{minted}

Дальнейшие операции похожи на работу в SymPy. Определим символьную
переменную \mintinline{julia}{x} и возьмём синус от этой переменной.
\begin{minted}{julia}
x = sympy.Symbol("x")
y = sympy.sin(x)
\end{minted}

Вычислим теперь $\sin(\pi)$:
\begin{minted}{julia}
y.subs(x, pi)
\end{minted}
Результат будет иметь тип \mintinline{julia}{PyObject}. Чтобы его
использовать в дальнейшем, следует привести его к какому-либо
числовому типу. Например, следующим образом:
\begin{minted}{julia}
y.subs(x, pi) |> float
\end{minted}
Это пример, кроме всего прочего, демонстрирует функциональную природу
Julia. В данном выражении оператор \mintinline{julia}{|>} задаёт
действие функции справа (по аналогии с конвейером в оболочке Unix). То
есть записи \mintinline{julia}{f(x)} и \mintinline{julia}{x |> f}
эквивалентны.

Однако, вышеприведённая запись выглядит несколько тяжеловесной.
Для простейших операций требуется писать много дополнительного кода,
не несущего никакого содержания. К счастью, можно использовать
стандартные возможности Julia для расширений языка.
Для того, чтобы операции SymPy не выделялись на фоне остальных
вычислений Julia, вызов питоновской библиотеки производится с помощью
пакета SymPy.lj~\cite{sympy.lj:site}:
\begin{minted}{julia}
import Pkg
Pkg.add("SymPy")
using SymPy
\end{minted}

Тогда вышерассмотренный пример примет следующий вид:
\begin{minted}{julia}
x = Sym("x")
y = sin(x)
subs(y, x, pi) |> float
\end{minted}
Для объектов SymPy используется тип \mintinline{julia}{Sym}.

В принципе, на этом можно и завершить. Если читатель имел опыт работы
с SymPy в Python, он сможет работать и с SymPy.lj в Julia.

Язык Julia создавался таким образом, чтобы на нём легко могли писать
программы люди, перешедшие с других языков программирования с научной
и инженерной спецификой (таких, как Matlab, R, FORTRAN). По этой
причине Julia содержит много синтаксического сахара, в частности одну и ту же
операцию можно выполнить несколькими путями. Так, символьную
переменную можно задать с помощью конструктора:
\begin{minted}{julia}
x = Sym("x")
\end{minted}
Также можно задать с помощью макроса:
\begin{minted}{julia}
@syms x y z
\end{minted}
Также можно задать с помощью функции, имитирующей работу в SymPy из Python:
\begin{minted}{julia}
x, y = symbols("x y", commutative=false)
\end{minted}
Здесь, кроме всего, кроме объявления символических переменных
указываются и их свойства.

Следует заметить, что обязательное явное задание типа несколько
противоречит динамической природе Julia. Однако, в данном случае мы
используем внешнюю библиотеку, поэтому играем по чужим
правилам. Здесь можно провести аналогию с Python, который, при работе
с внешними библиотеками, использует соглашения этой
библиотеки, а не подменяет её синтаксис своим.

SymPy.lj даёт пользователю Julia возможность манипулировать
символьными выражениями, предлагает для этого удобный интерфейс.
Правда (учитывая производительность современных компьютеров) никакой
оптимизации по скорости выполнения не производится.

Приведём несколько примеров работы с SymPy.lj.

При операциях с алгебраическими выражениями весьма полезны операции
факторизации (\mintinline{julia}{factor}) и расширения (\mintinline{julia}{expand}).
\begin{minted}{julia}
x,y  = symbols("x, y")
factor(x^2 - 2x + 1)
\end{minted}

\begin{equation*}
  \left(x - 1\right)^{2}
\end{equation*}

\begin{minted}{julia}
expand((x-1)*(x+1))
\end{minted}
\begin{equation*}
  x^{2} - 1
\end{equation*}

В символьных выражениях можно применять конструкции Julia, например
включения\footnote{Включения (comprehensions) задают общий и
  эффективный способ построения массивов. Их синтаксис 
  похож на обозначения конструктора множеств в аксиоматике Цермело--Френкеля.}:
\begin{minted}{julia}
expand(prod([x-i for i in 1:5]))
\end{minted}
\begin{equation*}
  x^{5} - 15 x^{4} + 85 x^{3} - 225 x^{2} + 274 x - 120
\end{equation*}

Это происходит от того, что SymPy.lj представляет символьные матрицы
как массивы с элементами типа \mintinline{julia}{Sym}, например, как
\mintinline{julia}{Array{Sym}}.\footnote{Естественно, это влечёт за собой
дополнительные накладные расходы.}

Зададим символьную матрицу.
\begin{minted}{julia}
x, y = symbols("x y")
m = [x 1; 1 x]
\end{minted}
\begin{equation*}
  \left[ \begin{array}{rr}x&1\\1&x\end{array}\right]
\end{equation*}

Тогда с матрицами можно производить стандартные операции, например умножать:
\begin{minted}{julia}
m * m
\end{minted}
\begin{equation*}
  \left[ \begin{array}{rr}x^{2} + 1&2 x\\2 x&x^{2} + 1\end{array}\right]
\end{equation*}
\begin{minted}{julia}
m .* m
\end{minted}
\begin{equation*}
  \left[ \begin{array}{rr}x^{2}&1\\1&x^{2}\end{array}\right]
\end{equation*}

Также большинство стандартных операций с матрицами расширены для работы с
символьными значениями. Это, кстати, один из примеров реализации
множественной диспетчеризации.

Упомянутые выше функции \mintinline{julia}{factor} и
\mintinline{julia}{expand}
производят некоторое (необходимое нам) упрощение исходного выражения.
В SymPy есть много функций для выполнения различного рода упрощений.
Например, 
существует общая функция называемая \mintinline{julia}{simplify},
которая пытается получить простейшую форму выражения, используя для
этого разнообразные эвристики:
\mintinline{julia}{simplify}:
\begin{minted}{julia}
a = (x + x^2)/(x*sin(y)^2 + x*cos(y)^2)
simplify(a)
\end{minted}
\begin{equation*}
  x + 1
\end{equation*}

Для решения уравнений и систем уравнений можно использовать функцию
\mintinline{julia}{solve}. Например, решим уравнение
\begin{equation}
  x^2 + 3x + 2 = 0
\end{equation}
относительно $x$:
\begin{minted}{julia}
x  = symbols("x")
solve(x^2 + 3*x + 2, x)
\end{minted}
\begin{equation*}
  \left[ \begin{array}{r}-2\\-1\end{array} \right]
\end{equation*}

Для систем уравнений можно использовать векторную нотацию:
\begin{minted}{julia}
x,y  = symbols("x y")
eq1 = x + y - 1
eq2 = x - y - 2
solve([eq1, eq2], [x, y])
\end{minted}
\begin{minted}{julia}
Dict{Any,Any} with 2 entries:
  y => -1/2
  x => 3/2
\end{minted}

И наконец, дифференцирование и интегрирование.

Функция \mintinline{julia}{diff} используется для вычисления
производной символьных выражений. Можно вычислить частные производные
и производные высшего порядка. Зададим, например, функции:
\begin{minted}{julia}
x, y = symbols("x y")
f(x) = exp(-x) * sin(x)
g(x, y) = x^2 + 17*x*y^2
\end{minted}
Вычислим производную $\dv{f(x)}{x}$:
\begin{minted}{julia}
 diff(f(x))
\end{minted}
\begin{equation*}
  - e^{- x} \sin{\left(x \right)} + e^{- x}
  \cos{\left(x \right)}
\end{equation*}
Мы здесь опустили аргумент, по которому ведётся
дифференцирование. Впрочем, его можно и обозначать:
\begin{minted}{julia}
diff(f(x), x)
\end{minted}

Также можем получить производную более высокого порядка $\dv[3]{f(x)}{x}$:
\begin{minted}{julia}
diff(f(x), x, 3)
\end{minted}
\begin{equation*}
  2 \left(\sin{\left(x \right)} + \cos{\left(x \right)}\right) e^{- x}
\end{equation*}

Можно взять и частную производную $\pdv{g(x,y)}{x}$:
\begin{minted}{julia}
diff(g(x,y), x)
\end{minted}
\begin{equation*}
  2 x + 17 y^{2}
\end{equation*}

Аналогично можно проводить символьное интегрирование.
Вычислим $\int (x^2+ x + 2) \dd{x}$:
\begin{minted}{julia}
x,y = symbols("x y")
integrate(x^2 + x + 2)
\end{minted}
\begin{equation*}
  \frac{x^{3}}{3} + \frac{x^{2}}{2} + 2 x
\end{equation*}
Или двойной интеграл $\int \dd{y} \int \dd{x} xy$:
\begin{minted}{julia}
    integrate(x*y, (x, y))
\end{minted}
\begin{equation*}
  \frac{y^{3}}{2}
\end{equation*}
Можно вычислить и определённый интеграл $\int_{0}^{1} x \dd{x}$:
\begin{minted}{julia}
integrate(x^2, (x, 0, 1))
\end{minted}
\begin{equation*}
  \frac{1}{3}
\end{equation*}

Таким образом мы имеем очень богатую систему компьютерной алгебры. И
работать с ней можно прямо из среды Julia.
  \section{Symata --- язык компьютерной алгебры для Julia}
\label{sec:symata}

При том, что SymPy является великолепной системой компьютерной
алгебры, у неё есть небольшой недостаток. Это готовая система,
<<работающая из коробки>>. Она предоставляет <<пользовательский слой>>
для символьных вычислений, но не <<слой разработчика>>, позволяющий
писать программный код. Для разработки можно использовать язык
программирования для задач символьной математики
Symata~\cite{symata.jl:site}\footnote{В случае Symata язык <<слоя
  разработчика>> и язык <<пользовательского слоя>> совпадают. Поэтому
  Symata является одновременно и языком программирования, и системой
  компьютерной алгебры.}. Язык Symata реализован на языке Julia. В качестве
основы для дизайна языка Symata принят язык
Wolfram~\cite{wolfram:book:elementary-introduction}.

Для использования данной системы необходимо установить пакет Symata:
\begin{minted}{julia}
import Pkg
Pkg.add("Symata")
\end{minted}

Загружается Symata стандартным образом:
\begin{minted}{julia}
using Symata;
\end{minted}

Следует обратить внимание на следующее свойство Symata. Если SymPy.lj
проектировался таким образом, чтобы его можно было использовать
напрямую из среды программирования Julia, бесшовно, то Symata является
предметно-ориентированным языком, написанным на Julia. Он достаточно
сильно отличается от языка Julia. Поэтому необходимо явно
переключаться между этими языками. Переключение в режим языка Julia
осуществляется с помощью следующей функции
\begin{minted}{julia}
Julia()
\end{minted}

Обратно в режим Symata можно переключиться с помощью функции
\begin{minted}{julia}
isymata()
\end{minted}

Вычисления в Symata похожи на работу в системе Mathematica. Приведём
пример. Пусть нам надо вычислить значение косинуса:
\begin{minted}{julia}
expr = Cos(pi * x)
\end{minted}
$$ \text{Cos} \!  \left(  \pi  \ x \right) $$

Зададим значение переменной $x$:
\begin{minted}{julia}
x = 1/3
\end{minted}
$$ \dfrac{1}{3} $$

Теперь собственно можно получить значение выражения \mintinline{julia}{expr}:
\begin{minted}{julia}
expr
\end{minted}
$$ \dfrac{1}{2} $$

Изменим значение переменной $x$:
\begin{minted}{julia}
x = 1/6
\end{minted}
$$ \dfrac{1}{6} $$

Тогда изменится и значение вычисляемого выражения:
\begin{minted}{julia}
expr
\end{minted}
$$ \dfrac{3^{\frac{1}{2}}}{2} $$

Сбросим значение переменной $x$. Тогда искомое выражение перестанет вычисляться:
\begin{minted}{julia}
Clear(x)
expr
\end{minted}
$$ \text{Cos} \!  \left(  \pi  \ x \right)  $$

Фактически средствами Julia реализован язык Wolfram. Так что,
наверное, нет необходимости более подробно освещать здесь синтаксис
системы Symata. Впрочем, стоит остановиться на некоторых отличиях
языка Symata от языка Wolfram.
Большинство этих отличий проистекает из того желания, чтобы Symata
наследовала синтаксис Julia. Например, комментарий задаётся символом
\mintinline{julia}{#}, в то время как в языке Wolfram комментарий
выглядит следующим образом:
\begin{minted}{mathematica}
(* comment *)
\end{minted}

Для задания списка используются не фигурные скобки
\mintinline{mathematica}{{ }}, а квадратные:
\begin{minted}{julia}
[a,b,c]
\end{minted}
$$  \left[ a,b,c \right]  $$
Элементы списка могут разделяться как запятыми, так и переходом на
новую строку:
\begin{minted}{julia}
[
    a
    c + d
    Expand((x+y)^2)
]
\end{minted}
$$  \left[ a,c + d,x^{2} + 2 \ x \ y + y^{2} \right]  $$

Аргумент функции задаётся не в квадратных скобках, как в Wolfram, а в
круглых, как в Julia:
\begin{minted}{julia}
f(x)
\end{minted}
$$ f \!  \left( x \right)  $$

Подобно функциям в Julia, некоторые функции приобрели инфиксную нотацию.

Функция \mintinline{julia}{Map(f,list)} приобретает инфиксную форму
\begin{minted}{julia}
f % list
\end{minted}
Функция \mintinline{julia}{Apply(x,y)}:
\begin{minted}{julia}
x .%  y
\end{minted}

Более подробно можно посмотреть в документации к Symata.

Можно сделать вывод, что на основе Julia (с использованием её мощных
возможностей по созданию предметно-ориентированных языков) реализован
исключительно удобный язык для компьютерной алгебры. При этом
интересно отметить, что для многих аналитических вычислений в языке
Symata используется библиотека SymPy.

\section{Компьютерная алгебра в Julia  для задач математического
  моделирования}
\label{sec:mtk}

В предыдущих разделах мы рассмотрели реализацию систем компьютерной
алгебры общего назначения на языке Julia. Но также большой интерес
вызывают специализированные, предметно-ориентированные применения
компьютерной алгебры.

Пакет ModelingToolkit.jl~\cite{modelingtoolkit.jl:site} предлагает специализированный язык
компьютерной алгебры для задач математического моделирования. Для
этого он использует возможности метапрограммирования~\cite{lammel:2018:software-languages} языка Julia.
Отметим, что основным режимом работы для ModelingToolkit.jl является
не интерактивный режим, а пакетный.

Для задания символьных переменных служит макрос
\mintinline{julia}{@variables}:
\begin{minted}{julia}
@variables x y
\end{minted}

После этого данные символьные переменные можно использовать в
символьных выражениях (сохраняя синтаксис Julia):
\begin{minted}{julia}
z = x^2 + y
\end{minted}

Для моделирования непрерывных динамических систем необходимо
использовать производные. В пакете ModelingToolkit.jl дифференциальные
операторы строятся с помощью макроса \mintinline{julia}{@derivatives}:
\begin{minted}{julia}
@variables t
@derivatives D'~t
\end{minted}
Здесь задан дифференциальный оператор $D = \dv{}{t}$. Количество
штрихов \mintinline{julia}{'} указывает на порядок дифференциального
оператора. Можно записать выражение:
\begin{minted}{julia}
z = t + t^2
D(z)
\end{minted}
Оператор дифференцирования не вычисляет ничего непосредственно в этот
момент, поскольку является <<ленивым>> оператором. Впрочем, мы можем
получить результат сразу, применив функцию \mintinline{julia}{expand_derivatives}:
\begin{minted}{julia}
expand_derivatives(D(z))
\end{minted}
\begin{equation*}
1 + 2t
\end{equation*}

Поскольку в Julia функции являются объектами первого порядка, то
объявленные символьные переменные являются, фактически,
функциями. Можно при объявлении переменных явно указать зависимости:
\begin{minted}{julia}
@variables t x(t) y(t)
\end{minted}

Эта зависимость учитывается при дифференцировании:
\begin{minted}{julia}
z = x + y*t
expand_derivatives(D(z))
\end{minted}
Последнее выражение будет раскрыто как
\begin{minted}{julia}
derivative(x(t), t) + y(t) + derivative(y(t), t) * t
\end{minted}

В качестве примера посмотрим, как решается классическая задача типа <<хищник--жертва>>~\cite{lotka:1925:physical-biology,kulyabov:2018:formalizms-stochastization}:
\begin{equation}
  \label{eq:lotka}
  \begin{dcases}
    \dv{x}{t} = ax - b y x,\\
    \dv{y}{t} = c x y - d y,
  \end{dcases}
\end{equation}
где $x$ --- количество жертв, $y$ --- количество хищников, $t$ ---
время, $a , b , c, d$ --- коэффициенты взаимодействия между видами.

\begin{figure}
  \centering
  \includegraphics[width=\linewidth]{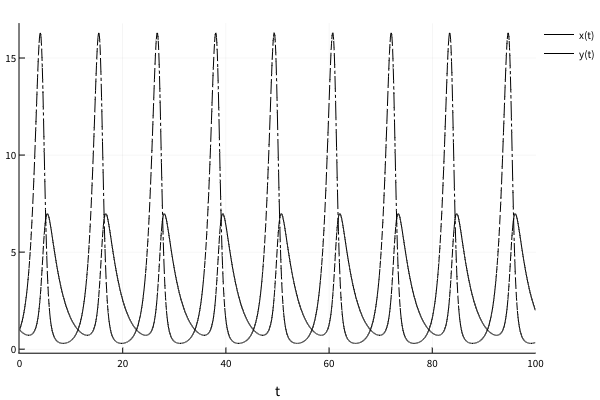}
  \caption{Решение системы~\eqref{eq:lotka}}
  \label{fig:lotka}
\end{figure}

Сначала загрузим необходимые пакеты (при необходимости):
\begin{minted}{julia}
import Pkg
Pkg.add("ModelingToolkit")
Pkg.add("OrdinaryDiffEq")
\end{minted}

Также загрузим пакет для поддержки построения графиков:
\begin{minted}{julia}
Pkg.add("Plots")
\end{minted}
Теперь подключим необходимые пакеты:
\begin{minted}{julia}
using ModelingToolkit
using OrdinaryDiffEq
using Plots
\end{minted}

Собственно, теперь можно решать нашу задачу. Определим переменные и
операторы дифференцирования. Часть переменных определим как параметры.
\begin{minted}{julia}
@parameters t a b c d
@variables x(t) y(t)
@derivatives D'~t
\end{minted}

Теперь просто перепишем исследуемую систему~\eqref{eq:lotka},
используя синтаксис Julia:
\begin{minted}{julia}
eqs = [D(x) ~ a *x - b*x*y,
       D(y) ~ c*x*y - d*y]
\end{minted}

Теперь применим символьные преобразования и приведём нашу систему к
виду, необходимому для пакета OrdinaryDiffEq.lj~\cite{rackauckas:2017:differentialequations.jl}:
\begin{minted}{julia}
sys = ODESystem(eqs)
\end{minted}

Дальнейшие манипуляции проводятся в рамках пакета
OrdinaryDiffEq.lj. Зададим начальные значения переменных:
\begin{minted}{julia}
u0 = [x => 1.0
      y => 1.0]
\end{minted}

Также зададим параметры задачи:
\begin{minted}{julia}
p  = [a => 1.1
      b => 0.4
      c => 0.1
      d => 0.4]
\end{minted}

Будем решать задачу на следующем временном промежутке:
\begin{minted}{julia}
tspan = (0.0,100.0)
\end{minted}

Создадим структуру, содержащую всю нашу задачу:
\begin{minted}{julia}
prob = ODEProblem(sys,u0,tspan,p; jac=true)
\end{minted}
Дополнительный параметр \mintinline{julia}{jac=true} указывает
системе символически сгенерировать оптимизированную функцию Якоби для
улучшения работы решателей дифференциальных уравнений.

И, наконец, применим решатель из пакета OrdinaryDiffEq.lj:
\begin{minted}{julia}
sol = solve(prob)
\end{minted}

Нарисуем график, представленный на рис.~\ref{fig:lotka}:
\begin{minted}{julia}
plot(sol)
\end{minted}

Рассмотренный подход очень хорошо вписывается в идеологию
специализированного языка для научных и инженерных расчётов. При
данном подходе для символьных вычислений используется не специальный
универсальный язык, а для каждого направления исследований создаётся свой
проблемно-ориентированный язык символьных расчётов. Это язык должен
бесшовно стыковаться с базовым языком, в нашем случае, с Julia.
\section{Заключение}
\label{sec:conclusion}

Наш обзор (далеко не полный) систем компьютерной алгебры для языка
Julia показал, что и сам язык, и его инфраструктура являются
достаточно зрелыми. При этом системы компьютерной алгебры языка Julia
охватывают широкий спектр применения: пользовательские системы
компьютерной алгебры (SymPy.lj), мощные языки компьютерной алгебры
(Symata.lj), проблемно-ориентированные языки компьютерной алгебры
(ModelingToolkit.jl). Всё это позволяет надеяться, что популярность
языка Julia будет возрастать не только в области численных, но и
символьных расчётов.

\def\acknowledgmentsname{Благодарности}

\begin{acknowledgments}

Публикация подготовлена при поддержке Программы РУДН «5-100»
и при финансовой поддержке РФФИ в рамках научного проекта
\textnumero~19-01-00645.

\end{acknowledgments}

  \ifthenelse{\boolean{xetex} \OR \boolean{luatex}}
  {\bibliographystyle{ugost2008l}}
  {\bibliographystyle{gost2008l}}

\bibliography{bib/julia-cas/cite}